\documentclass[twocolumn,showpacs,pra,aps]{revtex4}
\usepackage{array}
\usepackage{amsmath}
\usepackage{graphicx}
\usepackage{amstext}
\usepackage{amsfonts}
\begin{document}

\title{$GHZ$-type and $W$-type entangled coherent states: generation and Bell-type
inequality tests without photon counting}

\author{Hyunseok Jeong$^1$ and Nguyen Ba An$^{2,3}$}

\affiliation{\ $^{1}$Centre for Quantum Computer Technology, Department of
Physics, University of Queensland, Brisbane, Qld 4072, Australia
\\
$^{2}$Institute of Physics and Electronics, 10 Dao Tan, Thu Le, Ba Dinh, Hanoi, Vietnam
\\
$^{3}$School of Computational
Sciences, Korea Institute for Advanced Study, 207-43 Cheongryangni 2-dong,
Dongdaemun-gu, Seoul 130-722, Korea.
}

\date{\today}

\begin{abstract}
We study $GHZ$-type and $W$-type three-mode entangled coherent states. Both
the types of entangled coherent states violate Mermin's version of the Bell
inequality with threshold photon detection (i.e., without photon counting). Such
an experiment can be performed using linear optics elements and threshold
detectors with significant Bell violations
for $GHZ$-type entangled coherent states. However, to
demonstrate Bell-type inequality violations for $W$-type entangled coherent
states, additional nonlinear interactions are needed. We also propose an
optical scheme to generate $W$-type entangled coherent states in
free-traveling optical fields. The required resources for the generation are
a single-photon source, a coherent state source, beam splitters,
phase shifters, photodetectors, and Kerr nonlinearities.
Our scheme does not necessarily
require strong Kerr nonlinear interactions, i.e., weak nonlinearities
can be used for the generation of the $W$-type entangled coherent
states.
Furthermore, it is also robust
against inefficiencies of the single-photon source and the photon
detectors.
\end{abstract}

\pacs{03.67.Mn, 42.50.Dv, 03.65.Ud, 42.50.-p}

\maketitle
\section{Introduction}

Quantum entanglement is a resource for quantum information processing 
and quantum computing. A large class of entangled states violate
Bell-type inequalities \cite{Bell,CHSH,CH}, which means that the existence
of such states cannot be explained by any local theory. It has been known
that there exist at least two different types of multipartite
entanglement, namely, the $GHZ$-type \cite{GHZ} entanglement and 
the $W$-type entanglement \cite{W}. These two different types of
entanglement are not equivalent and cannot be converted to each other by
local unitary operations combined with classical communication \cite{W}.

Recently, entangled coherent states (ECSs) \cite{Sanders92} in
free-traveling optical fields have been found useful to perform tasks
such as quantum teleportation \cite{Enk,NBA}, quantum computation \cite
{JK,Ralph02,Ralph03}, entanglement purification \cite{JKpuri}, quantum error
corrections \cite{Glancy}, etc. Most of these schemes use 
single-mode coherent-state superpositions (CSSs) as qubits and two-mode
ECSs as quantum channels. Recently, Nguyen studied an optimal
quantum information processing via multi-mode $W$-type ECSs \cite{NBA}. In
particular, it was shown that there exists a quantum information protocol
which can be done only with $W$-type ECSs while $GHZ$-type ECSs fail to
accomplish such a task \cite{NBA}. As an example, the remote symmetric
entangling, that allows two distant parties to share a symmetric
entangled state, strictly requires $W$-type three-mode ECSs. Even
though there have been a number of studies on ECSs revealing their quantum
nonlocality \cite{ecsbell1,ecsbell2,Wilson,jeongbell} and usefulness for
quantum information processing \cite
{Enk,NBA,JK,Ralph02,Ralph03,JKpuri,Glancy}, most of them have focused
on two-mode ECSs while the characteristics of multi-mode ECSs,
particularly the $W$-type ECSs, have been relatively less known.

A $GHZ$-type ECS can be generated using beam splitters with a single-mode
CSS. Recently, there have been remarkable theoretical suggestions which are
within reach of current technology \cite
{Martini,Dakna2,Mauro03,Lund04,Jeong04,Jeong05,Wang,KimPaternostro05} and an
experimental attempt \cite{Wenger} for the generation of single-mode CSSs in
free-traveling optical fields. However, the generation of $W$-type ECSs in
free-traveling optical fields is not straightforward from single-mode CSSs.
Very recently, Yuan \textit{et al.}, suggested a scheme to generate
$GHZ$-type and $W$-type ECSs in cavity fields \cite{chgroup}. However,
it should be noted that for most of tasks for quantum information processing \cite
{Enk,NBA,JK,Ralph02,Ralph03,JKpuri,Glancy,ecsbell1,ecsbell2,Wilson,jeongbell},
one needs to generate such ECSs \textit{in free-traveling optical fields}.

In this paper, we suggest a scheme to generate W-type ECSs in
free-traveling optical fields and study violations of the Mermin's version 
\cite{Mermin} of the Bell inequality \cite{Bell} for both the $GHZ$-type
and $W$-type ECSs. Our study is closely associated
with currently feasible experimental elements in quantum optics. The Bell's
inequality test can be performed using linear optics elements and threshold
detectors with large violations for $GHZ$-type ECSs but it requires
additional nonlinear elements for $W$-type ECSs. As for our
generation scheme of $W$-type ECSs, it requires a 
single-photon source, a coherent state source, beam splitters, phase
shifters and threshold photodetectors as well as weak
nonlinearities. Our generation scheme is robust against
inefficiencies of the single-photon source and the photodetectors.

We organize our paper as follows. In section II we first outline possible schemes to generate multi-mode GHZ-type ECSs and then test the Bell-Mermin inequality for the three-mode case using photon parity (subsection II A) and photon threshold (subsection II B) measurements. Section III also has two subsections. Subsection III A presents a mechanism of generation of three-mode W-type ECS whose violation of the Bell-Mermin inequality is examined in subsection III B by means of photon threshold measurements. Finally, we conclude in section IV.

\section{$GHZ$-type entangled coherent states}

An $N$-mode $GHZ$-type ECS is defined as 
\begin{equation}
|GHZ,\alpha \rangle =c_{1}|\alpha ,\alpha ,...,\alpha \rangle
_{1...N}+c_{2}|-\alpha ,-\alpha ,...,-\alpha \rangle _{1...N}  \label{ghz}
\end{equation}
where $|\alpha ,\alpha ,...,\alpha \rangle _{1...N}=|\alpha \rangle
_{1}|\alpha \rangle _{2}...|\alpha \rangle _{N}$ with $|\alpha \rangle _{i}$
a coherent state of amplitude $\alpha $ while complex
coefficients $c_{1}$ and $c_{2}$ should satisfy the normalization
condition. Such an entangled state can be generated with a
single-mode CSS 
\begin{equation}
|CSS\rangle =c_{1}|\alpha \rangle +c_{2}|-\alpha \rangle  \label{css}
\end{equation}
and beam splitters. Recently, several feasible suggestions have been made
for the generation of CSSs in free-traveling optical fields \cite
{Martini,Dakna2,Mauro03,Lund04,Jeong04,Jeong05,Wang,KimPaternostro05}. For
example, it was found that simply squeezing a single photon results in a
very good approximation of a single-mode CSS with amplitude $\alpha
\leq 1.2$ \cite{Lund04}. It was also pointed out that a weak Kerr
nonlinearity can be useful to generate CSSs even under realistic decoherence 
\cite{Jeong05}. The beam splitter operator ${\hat{B}}(r,\phi )$ acting on
two arbitrary modes $a$ and $b$ is represented as 
\begin{equation}
\hat{B}(r,\phi )=\exp \{\frac{\theta }{2}
(e^{i\phi }\hat{a}^{\dagger }\hat{b}
-e^{-i\phi }\hat{b}^{\dagger }\hat{a})\}.
\end{equation}
where the reflectivity $r$ and transmittivity $t$
are determined by $\theta $ as 
\begin{equation}
r=\sin \frac{\theta }{2},~~~t=\cos \frac{\theta }{2},
\end{equation}
and $\phi $ specifies the phase difference between the reflected
and transmitted fields.
We assume $\phi=\pi$ throughout this paper.  
 In particular, if a three-mode $GHZ$-type
ECS of the form (\ref{ghz}), where $c_{1}=-c_{2}$, needs to be generated,
one can pass the CSS,
$|\sqrt{3}\alpha \rangle -|-\sqrt{3}\alpha \rangle$
(unnormalized), through two beam splitters BS1 with
$r_{1}=1/\sqrt{3}$ and BS2 with $r_{2}=1/\sqrt{2}$
successively as shown in Fig.~1. In general, to
generate a $GHZ$-type ECS with an arbitrary mode number $N$, one should 
pass a CSS of the form
$|\sqrt{N}\alpha \rangle \pm |-\sqrt{N}\alpha \rangle $
through a sequence of $N-1$ beam
splitters with $r_{1}=1/\sqrt{N},$ $r_{2}=1/\sqrt{N-1},$ ... and
$r_{N-1}=1/\sqrt{2}$ or, one can exploit a ``tree-scheme'' that uses only
50:50 beam splitters combined, if necessary, with state-parity measurements
(see, e.g. Ref.~\cite{NguyenKim}).

\begin{figure}[tbp]
\centerline{\scalebox{0.9}{\includegraphics{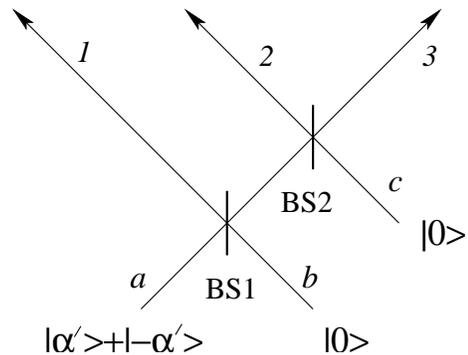}}}
\caption{An example of the generation of a $GHZ$-type three-mode ECS in
free-traveling fields from a CSS using two beam splitters, BS1 with 
$r_{1}=1/\protect\sqrt{3}$ and BS2 with 
$r_{2}=1/\protect\sqrt{2}$, where $r_{1}$ and 
$r_{2}$ are reflectivities of the beam splitters.
Note that $\alpha ^{\prime}=\protect\sqrt{3}\alpha $.}
\label{fig:wbell}
\end{figure}

\subsection{Bell-inequality violations of $GHZ$-type ECSs using photon
parity measurements}

We now study the Bell-Mermin inequality for a three-mode $GHZ$-type ECS.
Bell inequality violations for two-mode ECSs have been studied with photon
parity and photon threshold measurements \cite{Wilson,jeongbell}. We first
study the Bell-Mermin inequality based upon the parity measurement and the
displacement operation \cite{BW}. For that purpose we define an
observable $\Pi (\beta )$ as 
\begin{equation}
\Pi (\beta )=D(\beta )\sum_{n=0}^{\infty }\Big(|2n\rangle \langle
2n|-|2n+1\rangle \langle 2n+1|\Big)D^{\dagger }(\beta )  \label{eq:bw}
\end{equation}
where $D(\beta )$ is the displacement operator, 
$D(\beta )=\exp [\beta \hat{a}^{\dagger }-\beta ^{*}\hat{a}]$,
for bosonic operators $\hat{a}$ and $\hat{a}^{\dag }$.
Note that the eigenvalue of the observable $\Pi (\beta )$ is 1
when an even number of photons is detected and it is $-1$ when an odd number
of photons is detected. The Bell-Mermin inequality \cite{Bell,Mermin} based
on the observable $\Pi (\beta )$ is 
\begin{eqnarray}
&&BM_{\Pi }=|\langle \Pi (\beta _{1})
\Pi (\beta _{2})\Pi (\beta _{3})\rangle
-\langle \Pi (\beta _{1})
\Pi (\beta _{2}^{\prime })\Pi (\beta _{3}^{\prime
})\rangle   \nonumber \\
&&~~~~~-\langle \Pi (\beta _{1}^{\prime })
\Pi (\beta _{2})\Pi (\beta _{3}^{\prime
})\rangle -\langle \Pi (\beta _{1}^{\prime })
\Pi (\beta _{2}^{\prime })\Pi
(\beta _{3})\rangle |\leq 2,  \nonumber \\ \label{ffff-a}
\end{eqnarray}
where $\Pi (\beta _{1})\Pi (\beta _{2})\Pi (\beta _{3})=\Pi _{1}(\beta
_{1})\otimes \Pi _{2}(\beta _{2})\otimes \Pi _{3}(\beta _{3})$ and we shall
call $BM_{\Pi }$ the Bell-Mermin function. The average value, $\langle \Pi
(\beta _{1})\Pi (\beta _{2})\Pi (\beta _{3})\rangle $, can be calculated
using the identity 
\begin{equation}
\langle \Pi (\beta _{1})\Pi (\beta _{2})
\Pi (\beta _{3})\rangle =\frac{\pi^3}{8}
W(\beta _{1},\beta _{2},\beta _{3})
\end{equation}
where $W(\beta _{1},\beta _{2},\beta _{3})$ represents the Wigner function
of the state of interest. The Wigner function of the three-mode 
$GHZ$-type ECS in Eq.~(\ref{ghz}), for the case of $c_{1}=\pm c_{2}$, can be
calculated as follows \cite{Walls}.
First, the characteristic function $\chi_\pm(\eta_1,
\eta_2,\eta_3)$
can be obtained as
\begin{equation}
\chi_\pm(\eta_1,\eta_2,\eta_3)
={\rm Tr}[\rho_{GHZ}D(\eta_1) D(\eta_2) D(\eta_3)]
\label{eq:crf}
\end{equation}
where 
$\rho_{GHZ}=|GHZ,\alpha \rangle \langle GHZ,\alpha|$ and
$D(\eta_1)D(\eta_2) D(\eta_3)\equiv
 D_1(\eta_1)\otimes D_2(\eta_2)\otimes D_3(\eta_3)$. 
The subscript plus (minus) sign of the characteristic function in Eq.~(\ref{eq:crf})
denotes $c_{1}=+c_{2}$ ($c_{1}=-c_{2}$).
The characteristic function is then
\begin{widetext}
\begin{equation}
\begin{aligned}
\chi_\pm(\eta_1,\eta_2,\eta_3)
=&\langle GHZ,\alpha| D(\eta_1)D(\eta_2) D(\eta_3)|GHZ,\alpha \rangle\\
=&
M_\pm
e^{-\frac{1}{2}|\eta_1|^2
-\frac{1}{2}|\eta_2|^2
-\frac{1}{2}|\eta_3|^2}
\Big(\exp[\eta_1\alpha^*-\eta_1^*\alpha
+\eta_2\alpha^*-\eta_2^*\alpha
+\eta_3\alpha^*-\eta_3^*\alpha]\\
&~~~+\exp[-\eta_1\alpha^*+\eta_1^*\alpha
-\eta_2\alpha^*+\eta_2^*\alpha
-\eta_3\alpha^*+\eta_3^*\alpha
]\\
&~~~\pm\exp[-6|\alpha|^2-\eta_1\alpha^*-\eta_1^*\alpha
-\eta_2\alpha^*-\eta_2^*\alpha
-\eta_3\alpha^*-\eta_3^*\alpha]\\
&~~~\pm\exp[-6|\alpha|^2+\eta_1\alpha^*+\eta_1^*\alpha
+\eta_2\alpha^*+\eta_2^*\alpha
+\eta_3\alpha^*+\eta_3^*\alpha]
\Big)
\end{aligned}
\end{equation}
%\end{widetext}
where $M_\pm=1/(2\pm 2e^{-6|\alpha |^{2}})$.
The three-mode Wigner function is obtained from the characteristic function
as
\begin{equation}
W_\pm(\beta_1,\beta_2,\beta_3)
=\frac{1}{\pi^6}
\int d^2\eta_1 d^2\eta_2 d^2\eta_3
\exp[\eta_1^*\beta_1-\eta_1\beta_1^*
+\eta_2^*\beta_2-\eta_2\beta_2^*
+\eta_3^*\beta_3-\eta_3\beta_3^*
]\chi_\pm(\eta_1,\eta_2,\eta_3).
\end{equation}
After the integration, the Wigner function of the $GHZ$-type ECS is
%\begin{widetext} 
\begin{eqnarray}
W_\pm(\beta_1,\beta_2,\beta_3)=N_\pm
\bigg\{\exp[-2|\beta_1-\alpha|^2-2|\beta_2
-\alpha|^2-2|\beta_3-\alpha|^2]
~~~~~~~~~~~~~~~~~~~~\nonumber\\
+\exp[-2|\beta_1+\alpha|^2-2|\beta_2+\alpha|^2
-2|\beta_3+\alpha|^2]
~~~~~~~~~~~~~~~~~~\nonumber\\
\pm e^{-6|\alpha|^2}
\Big(\exp[-2(\beta_1-\alpha)(\beta_1+\alpha)^*
-2(\beta_2-\alpha)(\beta_2+\alpha)^*
-2(\beta_3-\alpha)(\beta_3+\alpha)^*]\nonumber\\
+\exp[-2(\beta_1-\alpha)^*(\beta_1+\alpha)
-2(\beta_2-\alpha)^*(\beta_2+\alpha)
-2(\beta_3-\alpha)^*(\beta_3+\alpha)]\Big)
\bigg\}
\end{eqnarray}
\end{widetext}
with $N_{\pm }=8/[\pi ^{3}(2\pm 2e^{-6|\alpha |^{2}})]$.
\begin{figure}[tbp]
\centerline{(a)} \centerline{\scalebox{0.8}{\includegraphics{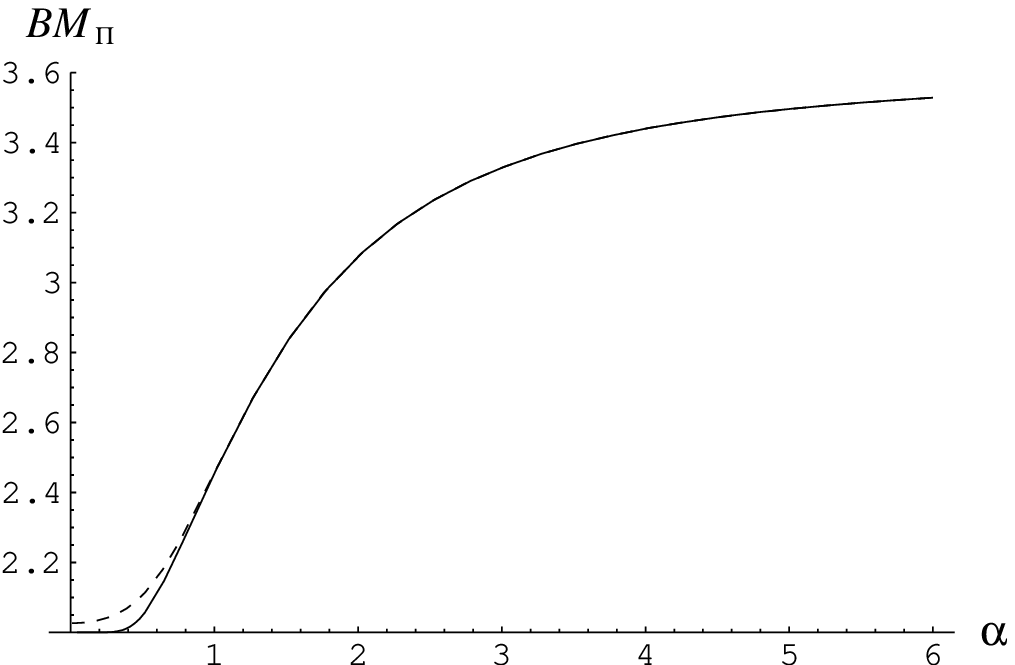}}} 
\centerline{(b)} \centerline{\scalebox{0.8}{\includegraphics{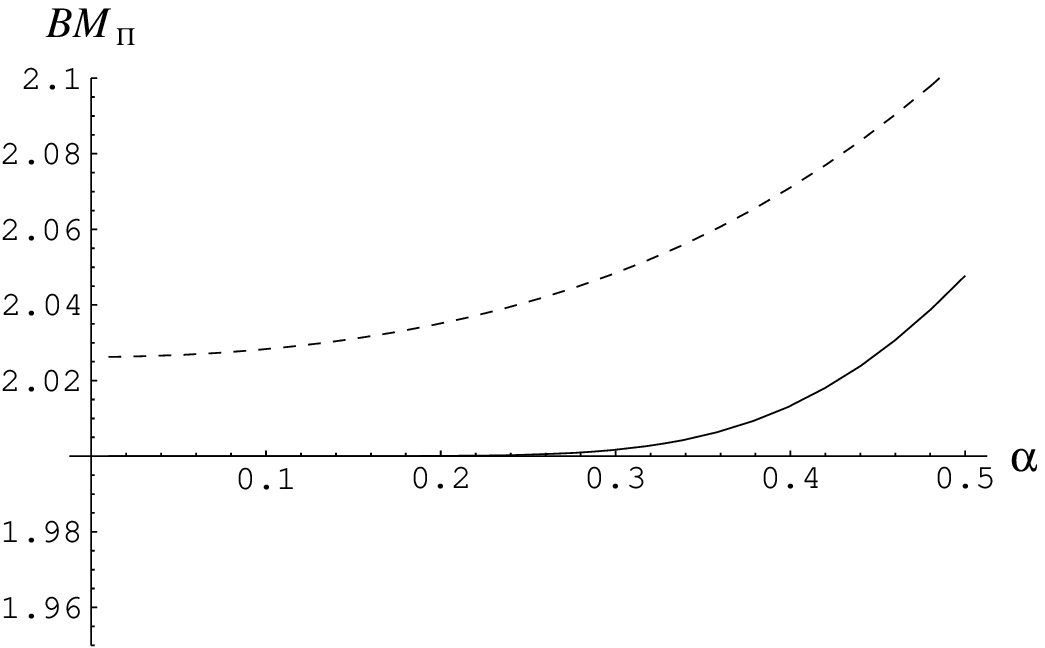}}}
\caption{ Violations of Bell-Mermin inequality for the $GHZ$-type ECS of
amplitude $\alpha$ using parity measurements. $BM_\Pi$ represents the
Bell-Mermin function defined in the text which is supposed to be not greater
than 2 by a local hidden variable theory. (a) As $\alpha$ increases the
violation increases and saturates at $\approx 3.6$ for the $GHZ$-type ECSs
of $c_1=c_2$ (solid line) and $c_1=-c_2$ (dashed line). (b) Even when 
$\alpha$ is extremely small the $GHZ$-type ECS of $c_1=-c_2$ (dashed line)
still violates the Bell-Mermin inequality while the $GHZ$-type ECS of 
$c_1=c_2$ does not.}
\label{fig:parity}
\end{figure}
 The
Bell-Mermin function $BM_{\Pi }$ has 12 variables and it is highly
nontrivial to find the global maximum values of $BM_{\Pi }$ for all the 12
variables. Fortunately, some local maximum values which violate the
Bell-Mermin inequality can be found numerically using the method of steepest
descent \cite{nume}, which has been plotted in Fig.~\ref{fig:parity}.

Fig.~\ref{fig:parity} shows that as $\alpha $ increases the value 
$BM_{\Pi }$ of the Bell-violation first grows and then
asymptotically approaches the value of $3.6$ when
$\alpha\rightarrow \infty$
for the $GHZ$-type ECSs of $c_{1}=c_{2}$ and 
$c_{1}=-c_{2}$.
For instance, when $\alpha =10$, the Bell-Mermin
function is found to be $BM_{\Pi }\approx 3.58$ at
$Re[\beta _{1}]=Re[\beta_{2}]=Im[\beta _{3}]
=Re[\beta _{1}^{\prime }]=Re[\beta _{2}^{\prime}]
=Re[\beta _{3}^{\prime }]=0,$ $Im[\beta _{1}]
=-Im[\beta _{1}^{\prime}]=-0.020,$
$Im[\beta _{2}]=-Re[\beta _{3}]=-0.0519$
and $Im[\beta_{2}^{\prime }]
=-Re[\beta _{3}^{\prime }]=-0.0259$
for both the cases of $c_{1}=c_{2}$ and $c_{1}=-c_{2}$.
It is interesting to note that even when 
$\alpha $ is extremely small, the $GHZ$-type ECS of
$c_{1}=-c_{2}$ (dashed curve in Fig.~2b)
still violates the Bell-Mermin inequality 
whereas the $GHZ$-type ECS of $c_{1}=c_{2}$
does not (solid curve in Fig. 2 b).
This is due to the singular behavior of the $GHZ$-type ECS of 
$c_{1}=-c_{2}$ when $\alpha $ approaches zero. When $\alpha $ approaches
zero, the $GHZ$-type ECS becomes a vacuum product state without
entanglement, i.e. $|0\rangle |0\rangle |0\rangle $, unless $c_{1}=-c_{2}$.
However, if $c_{1}=-c_{2}$, as $\alpha $ approaches zero, the $GHZ$-type ECS
generated using beam splitters as shown in Fig.~1 will approach 
\begin{equation}
\frac{1}{\sqrt{3}}(|1\rangle |0\rangle |0\rangle +|0\rangle
|1\rangle |0\rangle +|0\rangle |0\rangle |1\rangle)
\label{111}
\end{equation}
which is a highly nonlocal entangled state. This can be understood
as follows. In order to generate the $GHZ$-type ECS of $c_{1}=-c_{2}$ using
two beam splitters, BS1 and BS2 as sketched in Fig.~1, a CSS of 
$c_{1}=-c_{2}$ is needed. The CSS of $c_{1}=-c_{2}$ approaches the 
single-photon state, $|1\rangle $, for $\alpha \rightarrow 0$
\cite{jeongbell}. This single-photon state after passing through BS1 and
BS2 in Fig.~1 results in the state given by Eq.~(\ref{111}).
Therefore, one can expect that the $GHZ$-type ECS approaches the state 
(\ref{111}) showing Bell-type inequality violations only when 
$c_{1}=-c_{2}$.
Such a singular behavior is in agreement with the previous studies for
two-mode ECSs \cite{Wilson,jeongbell}. Namely, a two-mode ECS, 
$|\alpha ,\alpha \rangle -|-\alpha ,-\alpha \rangle$ (unnormalized),
violates the Bell inequality even in the limit
$\alpha \rightarrow 0$
because the two-mode ECS becomes a single photon entangled state,
$(|0,1\rangle +|1,0\rangle )/\sqrt{2}.$
Nevertheless, the two-mode
ECS, $|\alpha ,\alpha \rangle +|-\alpha ,-\alpha \rangle $
(unnormalized), ceases to violate the Bell inequality in the limit 
$\alpha \rightarrow 0$ as in this limit it becomes a vacuum product
state $|0\rangle |0\rangle $.

The displacement operation used in this type of Bell-inequality tests can be
effectively performed in an optics experiment by means of a beam
splitter with the transmission coefficient close to unity and a strong
coherent state being injected into the other input port \cite{displacement}.
However, the photon parity measurements should distinguish between odd and
even number of photons. This is extremely hard because the detectors which 
perfectly discriminate between neighbouring photon numbers
(i.e. $n$ and $n+1$ photons) do not exist at
the present status of technology.
On the other hand, the threshold photon
detection, which discriminate between presence and absence of photons, are 
available with a reasonably high probability using current
technology \cite{Takeuchi}. Therefore, in what follows, we shall focus on
the Bell-Mermin inequality test using only threshold photon detection and
displacement operations.

\subsection{Bell-inequality violations of $GHZ$-type ECSs using photon
threshold measurements}

\begin{figure}[tbp]
\centerline{\scalebox{0.8}{\includegraphics{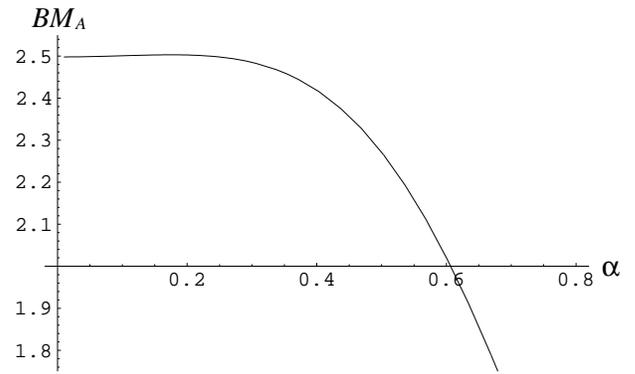}}}
\caption{ Violations of Bell-Mermin inequality for a $GHZ$-type ECS of
amplitude $\alpha$ using threshold photodetectors. $BM_A$ represents the
Bell-Mermin function defined in the text, Eq.~(\ref{ffff}), which is
supposed to be less than or equal to 2 by a local hidden variable theory. As 
$\alpha$ increases the violation first decreases and then ceases starting
from $\alpha\approx0.6$.}
\label{fig:ghzbell}
\end{figure}

We define an observable $A(\beta )$ for a Bell-Mermin inequality test with
the threshold photon measurements and the displacement operations as 
\begin{equation}
A(\beta )=D(\beta )^{\dagger }\big
(|0\rangle \langle 0|-\sum_{n=1}^{\infty}
|n\rangle \langle n|\big)D(\beta ),  \label{eq:op-A}
\end{equation}
whose eigenvalue is 1 when no photon are detected and $-1$
when any $n\geq 1$ photon(s) are detected. The Bell-Mermin
inequality based on the observable $A(\beta )$ reads 
\begin{eqnarray}
&&BM_{A}=|\langle A(\beta _{1})A(\beta _{2})
A(\beta _{3})\rangle -\langle
A(\beta _{1})A(\beta _{2}^{\prime })
A(\beta _{3}^{\prime })\rangle 
\nonumber \\
&&~~-\langle A(\beta _{1}^{\prime })
A(\beta _{2})A(\beta _{3}^{\prime})\rangle
-\langle A(\beta _{1}^{\prime })
A(\beta _{2}^{\prime })A(\beta
_{3})\rangle |\leq 2.~~~~~~ %\nonumber \\ 
\label{ffff}
\end{eqnarray}
In order to calculate $BM_{A}$ for the ECSs, we first compute the following
quantities: 
\begin{eqnarray}
&&J(\beta )\equiv \langle \alpha |A(\beta )
|\alpha \rangle =2\exp [-|\alpha
+\beta |^{2}]-1,  \label{eq:J} \\
&&K(\beta )\equiv \langle -\alpha |A(\beta )|-\alpha \rangle =2\exp
[-|\alpha -\beta |^{2}]-1, ~~~\\
&&L(\beta )\equiv \langle \alpha |A(\beta )|-\alpha \rangle\nonumber\\
&&~~~~~~=\exp [-\frac{1}{2}|\alpha +\beta |^{2}
-\frac{1}{2}|\alpha -\beta |^{2}
\nonumber\\
&&~~~~~~~~~~~~~~
+\alpha \beta ^{*}-\alpha ^{*}\beta ]\Big(2-e^{-(\alpha +\beta
)^{*}(\alpha -\beta )}\Big). %\nonumber\\
 \label{eq:L}
\end{eqnarray}
In terms of these quantities we have for the $GHZ$-type three-mode ECS 
\begin{eqnarray}
&&\langle A(\beta _{1})A(\beta _{2})A(\beta _{3})
\rangle=|c_{1}|^{2}J(\beta
_{1})J(\beta _{2})J(\beta _{3}) \nonumber\\
&&~~~~~~~~~~~~~~~~~+|c_{2}|^{2}K(\beta _{1})K(\beta _{2})K(\beta _{3})  
\nonumber\\
&&~~~~~~~~~~~~~~~~~+2\mathfrak{R}[c_{1}c_{2}^{*}L(\beta _{1})L(\beta _{2})L(\beta
_{3})]  \label{eq:ave}
\end{eqnarray}
from which one can calculate $BM_{A}$ in Eq.~(\ref{ffff}). The value of this
violation can be found numerically using again the method of
steepest descent \cite{nume}, which has been plotted for the $GHZ$-type ECSs
of $c_{1}=-c_{2}$ in Fig.~\ref{fig:ghzbell}. As can be seen from the figure,
the Bell-Mermin inequality is largely violated for a small $\alpha $. The
maximum value is found to be $BM_{A}\approx 2.5$ when
$\alpha\approx 0.18$ at $Re[\beta _{1}]=Re[\beta _{2}]
=\beta _{2}^{\prime}=\beta_{3}^{\prime}=0$,
$Im[\beta _{1}]=Im[\beta _{2}]=-0.371,$
$Re[\beta_{3}]=Im[\beta _{3}]=-0.295,$
and $Re[\beta _{1}^{\prime }]=Im[\beta_{1}^{\prime }]=0.173$. 
Even when $\alpha $ is extremely small, the inequality is still
significantly violated because we have considered the $GHZ$-type ECS of 
$c_{1}=-c_{2}$. As explained in the previous subsection, the $GHZ$-type ECS
with this particular relative phase approaches the single photon entangled
state (\ref{111}) for $\alpha \rightarrow 0$. However, the Bell-Mermin
inequality ceases to be violated when $\alpha \gtrsim 0.6$ as shown in 
Fig.~\ref{fig:ghzbell}. Comparing Figs.~\ref{fig:parity} and
\ref{fig:ghzbell}, we immediately learn that the photon threshold measurements
are more efficient to test the Bell-Mermin inequality for $GHZ$-type
ECSs with small amplitudes, while the photon parity measurements are
more efficient for $GHZ$-type ECSs with large amplitudes.

This result is due to the characteristic of the observable 
$A(\beta)$ in Eq.~(\ref{eq:op-A}) composed of the photon threshold
measurement and the displacement operations. It should be noted that
a Bell-type inequality test depends not only on the state being
considered but also on the type of the measurements and the type
of the random rotations used for the test. A previous study
\cite{jeongbell} presents a similar result for nonclassical
two-mode fields: two-mode ECSs and two-mode squeezed states
significantly violate the Bell-Clauser-Horne inequality \cite{CH}
using the photon threshold detection and the displacement operations only
when the coherent amplitudes (or the degree of squeezing) are small. All
these results suggest that the photon threshold measurements and the
displacement operations are efficient for a Bell-type inequality test for
the continuous variable states only when the average photon number
of the states is appropriately small. In this connection, the
photon threshold measurements and the displacement operations prove
to be very suitable as an efficient tool to test a Bell-type inequality in
a quantum optics experiment because $GHZ$-type ECSs with small amplitudes
are relatively easy to be generated from CSSs with small amplitudes
\cite{Lund04,Jeong05}.

\section{$W$-type entangled coherent states}

\subsection{Generation of $W$-type ECSs}

\begin{figure}[tbp]
\centerline{\scalebox{0.65}{\includegraphics{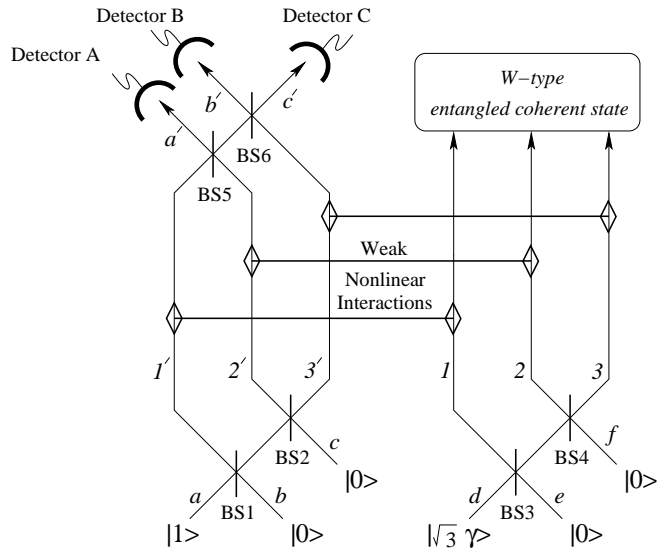}}}
\caption{A schematic of our scheme to generate a $W$-type ECS. A $W$-type
ECS is generated when detector B or detector C clicks. }
\label{fig:produce}
\end{figure}

A three-mode $W$-type ECS is \cite{NBA}
\begin{equation}
\begin{aligned}
&|W,\alpha \rangle =a_{1}|\alpha ,-\alpha ,-\alpha \rangle _{123}
+a_{2}|-\alpha ,\alpha ,-\alpha \rangle _{123}
\\
&~~~~~~~~~~~~~+a_{3}|-\alpha
,-\alpha ,\alpha \rangle _{123},  \label{eq:W}
\end{aligned}
\end{equation}
where the coefficients should meet the normalization condition. Our $W$-type
ECS generation scheme is depicted in Fig.~\ref{fig:produce}. We will first
describe how to generate a three-mode $W$-type ECS with 
$|a_{1}|=|a_{2}|=|a_{3}|$.
However, it can be simply generalized to
the case with arbitrary coefficients. In Fig.~\ref{fig:produce}, the
reflectivity of the first beam splitter, BS1, is $r_{1}=\sqrt{2/5}$
and that of the second beam splitter, BS2, is $r_2=\sqrt{2/3}$.
Through BS1 and BS2, the initial state
of the left part in Fig.~\ref{fig:produce} becomes 
\begin{equation}
\begin{aligned}
& \hat{B}_{BS2}\hat{B}_{BS1} |1,0,0\rangle_{abc}
=\frac{1}{\sqrt
5}\big(\sqrt{2}|1,0,0\rangle
+\sqrt{2}|0,1,0\rangle\\
&~~~~~~~~~~~~~~~~~~~~~~~~~~~~~~~~~~~~~~~~+ |0,0,1\rangle\big)_{1^\prime
2^\prime 3^\prime}
\end{aligned}
\end{equation}
On the right part of Fig.~4, three coherent states $|\gamma \rangle
_{1}|\gamma \rangle _{2}|\gamma \rangle _{3}$ are prepared using a coherent
state $|\sqrt{3}\gamma \rangle _{d}$ and two beam splitters BS3 and BS4 with
reflectivities $r_{3}=1/\sqrt{3}$ and $r_{4}=1/\sqrt{2}$, respectively. 
Then the cross Kerr nonlinear interactions are to be used
between mode $1^{\prime }$ ($2^{\prime }$, $3^{\prime }$) and mode 1 (2, 3)
as shown in Fig.~\ref{fig:produce}. The interaction Hamiltonian of the cross
Kerr nonlinear interaction between two arbitrary modes $a$ and $b$, 
$H_{CK}=\hbar \chi a^{\dagger }ab^{\dagger }b$, causes transformations
$|0\rangle _{a}|\gamma \rangle _{b}\rightarrow
|0\rangle _{a}|\gamma \rangle_{b}$ and 
$|1\rangle _{a}|\gamma \rangle _{b}\rightarrow |1\rangle
_{a}|\gamma e^{i\theta }\rangle _{b},$ where $\theta =\chi t$ with $\chi $
the coupling constant and $t$ the interaction time. Even though available
Kerr nonlinearities are extremely weak, very recently it has been discussed
that use of strong coherent fields may circumvent this problem even under
realistic assumption of decoherence \cite{Jeong05}. Therefore, in our
scheme, $\gamma $ is supposed to be large. The nonlinear interactions
transform the total state to
\begin{equation}
\begin{aligned}
& \frac{1}{\sqrt 5}\big( \sqrt{2}|1,0,0\rangle_{1^\prime
2^\prime 3^\prime}
|\gamma e^{i\theta},\gamma,\gamma\rangle_{123}\\
&~~~~~+\sqrt{2}|0,1,0\rangle_{1^\prime 2^\prime 3^\prime}
 |\gamma,\gamma
e^{i\theta},\gamma\rangle_{123}\\
&~~~~~+|0,0,1\rangle_{1^\prime 2^\prime
3^\prime} |\gamma,\gamma,\gamma e^{i\theta}\rangle_{123} \big).
\label{eq:wstep}
\end{aligned}
\end{equation}
Then, two 50:50 beam splitters, BS5 and BS6, are applied to
modes $1^{\prime}$, $2^{\prime }$ and $3^{\prime }$
as shown in Fig.~\ref{fig:produce}. It
is straightforward to verify that the total state after passing through BS5
and BS6 becomes 
\begin{widetext}
\begin{equation}
\frac{1}{\sqrt 5}|1,0,0\rangle_{a^\prime b^\prime c^\prime}
\big(|\gamma e^{i\theta},\gamma\rangle
+|\gamma,\gamma e^{i\theta}\rangle)|\gamma\rangle\big)_{123}
+\frac{1}{\sqrt{10}}
(|0,1,0\rangle+|0,0,1\rangle)_{a^\prime b^\prime c^\prime}
(|\gamma e^{i\theta},\gamma,\gamma\rangle
+|\gamma,\gamma e^{i\theta},\gamma\rangle
+|\gamma,\gamma,\gamma e^{i\theta}\rangle)_{123}.
\label{eq:wstep2}
\end{equation}
\end{widetext}
Now, detectors A, B and C (see Fig.~\ref{fig:produce}) are set to
detect photons of modes $a^{\prime }$, $b^{\prime }$ and $c^{\prime }$,
respectively. If detector A clicks the resulting state
(unnormalized) is reduced to 
\begin{equation}
(|\gamma e^{i\theta },\gamma \rangle 
+|\gamma ,\gamma e^{i\theta }\rangle
)_{12}|\gamma \rangle _{3}  \label{eq:wstep3}
\end{equation}
while either detector B or C clicks it is 
\begin{equation}
(|\gamma e^{i\theta },\gamma ,\gamma \rangle +|\gamma ,\gamma e^{i\theta
},\gamma \rangle +|\gamma ,\gamma ,\gamma e^{i\theta }\rangle )_{123}.
\label{eq:wstep4}
\end{equation}
The final step is to apply the displacement operators, $D(x)\otimes
D(x)\otimes D(x),$ with $x=-(\gamma +\gamma e^{i\theta })/2$, on
the state in (\ref{eq:wstep4}). Such operations are to change the state in 
(\ref{eq:wstep4}) to the symmetric form
in the phase space in Eq.~(\ref{eq:W}).
As we have pointed out previously, the displacement operation can be
effectively performed using a strong coherent fields (an additional local
oscillator in this case) and a biased beam  splitter.  It can be shown that
the final state after the displacement operations is 
\begin{equation}
N_{W}(|\alpha ,-\alpha ,-\alpha \rangle
+|-\alpha ,\alpha ,-\alpha \rangle
+|-\alpha ,-\alpha ,\alpha \rangle )_{123},  \label{eq:wstep5}
\end{equation}
where $\alpha =\gamma (e^{i\theta }-1)/2$ and $N_{W}=1/\sqrt{3+6e^{-4|\alpha
|^{2}}}$. However, the final displacement operations are only local unitary
transformations which cannot change entanglement nature of a quantum state.
Therefore, we stress that the generated state (\ref{eq:wstep4}) is
already a $W$-type entangled coherent state and the displacement operations 
mentioned above are just an optional step.

Note that the success probability of getting the $W$-type
three-mode ECS, Eq. (\ref{eq:wstep5}), is 3/5, while a two-mode
ECS, $\left( |\alpha ,-\alpha \rangle +|-\alpha ,\alpha
\rangle \right)_{12}$ (unnormalized), can be obtained with a
probability of 2/5 from the state in Eq.~(\ref{eq:wstep3}) using similar
displacement operations. Even though our case concerned 
$a_{1}=a_{2}=a_{3}$, arbitrary values of these coefficients can also be 
tailored by changing the reflectivities and phases of the beam
splitters BS1 and BS2. For example, if one needs to generate a $W$-type ECS
of $a_{1}=a_{2}=-a_{3}$, the phase of BS2, $\phi _{2}$, should be
set to be $\phi _{2}=\pi$. It is worth emphasizing that our scheme depicted
in Fig.~\ref{fig:produce} does not require highly efficient detectors
because the inefficiency of the detectors does not affect the quality of the 
generated $W$-type ECSs, yet it might decrease the
success probability to be lower than 3/5. Also of interest is the
fact that generally not all the three detectors are necessary in
our scheme. For example, only one of the
detectors B and C (but not both of them) would suffice to generate
a $W$-type ECS with a success probability of 3/10 when the detector
clicks. 
Furthermore, our scheme is robust to inefficiencies of the 
single-photon source as well as of the photon detectors. The
inefficiencies of the single-photon source and the photon detectors
will only reduce the success probability but will not affect the
quality of the $W$-type ECSs to be generated.

\subsection{Bell-inequality violations of $W$-type ECSs using photon
threshold measurements}

\begin{figure}[tbp]
\centerline{\scalebox{0.85}{\includegraphics{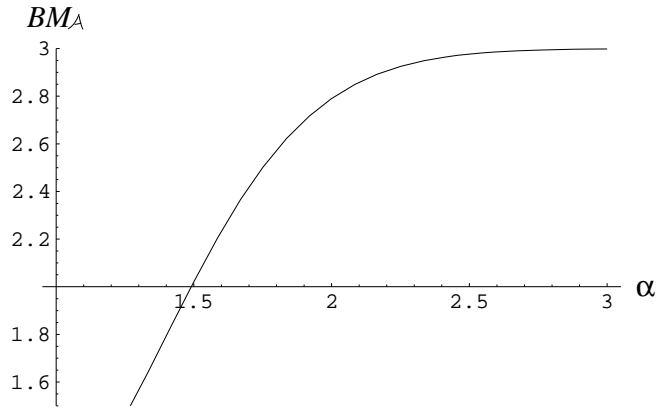}}}
\caption{ Violations of the Bell-Mermin inequality for a $W$-type ECS of
amplitude $\alpha$ using threshold photodetectors and nonlinear
interactions. The Bell-Mermin function $BM_{\mathcal{A}}$ is larger than the
classical bound $2$ for $\alpha\gtrsim 1.49$.}
\label{fig:wbell2}
\end{figure}

We now investigate the Bell-Mermin inequality for the $W$-type ECSs
in Eq.~(\ref{eq:wstep5}). According to our numerical study, the Bell-Mermin
inequality is not violated for $W$-type ECSs by the methods used
for $GHZ$-type ECSs in this paper. In this case, the required random
rotations for the Bell tests cannot be achieved by the displacement
operation \cite{Wu05}. Therefore, in this subsection, we shall 
alternatively approach the problem by treating the coherent state
$|\alpha \rangle $ and the vacuum $|0\rangle$ as a logical qubit basis
$\{|0_{L}\rangle ,|1_{L}\rangle\}$,
 namely, $|0_{L}\rangle \equiv |0\rangle $ and
 $|1_{L}\rangle \equiv |\alpha \rangle$.
 Since our approach here becomes a closer analogy of the ideal qubit case
for a $W$-type entangled state \cite{Cabello02} when $\alpha $ gets large,
it is expected that the Bell-Mermin inequality would strongly be
violated in this large-$\alpha $ limit. The first step to
make use of the logical qubits is to displace a $W$-type ECS of the form 
given by Eq.~(\ref{eq:wstep5}) to a $W$-type ECS of the following form
\begin{equation}
|\Psi_w\rangle=\mathcal{N}_{w}(|\alpha ,0,0\rangle
+|0,\alpha ,0\rangle +|0,0,\alpha
\rangle ),  \label{eq:a00}
\end{equation}
where $\mathcal{N}_{w}=1/\sqrt{3+6e^{-|\alpha |^{2}}}$. This transformation
can be done by the local displacement operations $D(\alpha ^{\prime
})\otimes D(\alpha ^{\prime })\otimes D(\alpha ^{\prime })$ on a $W$-type
ECSs $|W,\alpha ^{\prime }\rangle$ 
with $a_{1}=a_{2}=a_{3}$, where
$\alpha ^{\prime }=\alpha/2$.
One can directly transform the state
(\ref{eq:wstep4}) to the state (\ref{eq:a00})
by appropriate displacement operations. Let us introduce a unitary
transformation $U_{X}$ for a coherent-state qubit 
\begin{equation}
U_{X}=D(\frac{\alpha }{2})U_{K}(\pi/\chi )D(-\frac{\alpha }{2})  \label{???}
\end{equation}
where $U_{K}(t)=e^{iH_{K}t/\hbar }$, $H_{K}=\hbar \chi (a^{\dagger }a)^{2}$
with $t$ the interaction time in a single-mode Kerr nonlinear medium. In
other words, besides the displacement operations, additional single-mode
Kerr nonlinear interactions described by the interaction Hamiltonian $H_{K}$
with the interaction time $t=\pi /\chi $ need to be used for
necessary random rotations of a Bell-type inequality test. Even though such
strong nonlinear interactions are very demanding in a real experiment, there
was an experimental report for a successful measurement of giant Kerr
nonlinearity \cite{Hau}. Using the identity \cite{Yurke}, 
$U_{K}(\pi/\chi)|\alpha \rangle =e^{-i\pi /4}
(|\alpha \rangle +i|-\alpha \rangle )/\sqrt{2}$,
one can easily verify
\begin{eqnarray}
\begin{aligned}
U_X|0\rangle=\frac{e^{-i\pi/4}}{\sqrt{2}}(|0\rangle+i|\alpha\rangle),\\
U_X|\alpha\rangle=\frac{e^{-i\pi/4}}{\sqrt{2}}(i|0\rangle+|\alpha\rangle).
\label{eq:Kerr}
\end{aligned}
\end{eqnarray}
which corresponds to $-\pi /2$ rotation around the $x$ axis up to the
irrelevant global phase factor for the qubit basis $\{|0\rangle $, $|\alpha
\rangle \}.$ Because of the additional Kerr nonlinearities used, we define
an observable $\mathcal{A}(\tau)$, which is slightly different from $A(\beta )$
in Eq.~(\ref{eq:op-A}): 
\begin{equation}
\mathcal{A}(\tau)=U(\tau)^{\dagger }
\big(|0\rangle \langle 0|-\sum_{n=1}^{\infty
}|n\rangle \langle n|\big)U(\tau),  \label{eq:op-AA}
\end{equation}
where $\tau$ is either 0 or 1 and 
\begin{equation}
U(0)=\openone,~~U(1)=U_{X}.  \label{eq:op-UU}
\end{equation}
When $\alpha $ is large, the states $|0\rangle $ and $|\alpha \rangle $ are
eigenstates of the observable $\mathcal{A}(0)$ with the eigenvalues
$1$ and $-1$, respectively. The Bell-Mermin inequality is then constructed
as 
\begin{eqnarray}
\begin{aligned}
&&BM_{\cal A}=|\langle {\cal A}(\tau_1){\cal A}(\tau_2){\cal A}(\tau_3)\rangle
-\langle {\cal A}(\tau_1){\cal A}(\tau_2^\prime)
{\cal A}(\tau_3^\prime)\rangle~~~~~\\
&&-\langle {\cal A}(\tau_1^\prime){\cal A}(\tau_2)
{\cal A}(\tau_3^\prime)\rangle
-\langle {\cal A}(\tau_1^\prime){\cal A}(\tau_2^\prime)
{\cal A}(\tau_3)\rangle|\leq2.~~~
\label{eq:BMcala}
\end{aligned}
\end{eqnarray}
Taking $\tau_{1}=\tau_{2}=\tau_{3}=0$ and $\tau_{1}^{\prime }=\tau_{2}^{\prime
}=\tau_{3}^{\prime }=1$, and using Eqs.~(\ref{eq:a00}) to (\ref{eq:op-UU}),
it is straightforward to calculate 
\begin{eqnarray}
\langle \mathcal{A}(0)\mathcal{A}(0)\mathcal{A}(0)\rangle
=\frac{4 -e^{\alpha^2}}{2 + e^{\alpha^2}}\equiv {\cal V}(\alpha )
\end{eqnarray}
and 
\begin{eqnarray}
\begin{aligned}
\langle {\cal A}(0){\cal A}(1){\cal A}(1)\rangle
=\langle {\cal A}(1){\cal A}(0){\cal A}
(1)\rangle
=\langle {\cal A}(1){\cal A}(1){\cal A}(0)\rangle \\ 
=-\frac{2-2e^{-2\alpha^2}-7e^{-\alpha^2}-2e^{\alpha^2}}
{3(2+e^{\alpha^2})}
\equiv {\cal W}(\alpha)
\label{eq:r2}
\end{aligned}
\end{eqnarray}
where $\alpha$ was assumed to be real.
 Then the Bell-Mermin function is obtained as 
\begin{eqnarray}
BM_{\mathcal{A}}&=&|{\cal V}(\alpha )-3{\cal W}(\alpha )| \nonumber\\
&=&\Big|\frac{6-2e^{-2\alpha^2}-7e^{-\alpha^2}-3e^{\alpha^2}}
{2+e^{\alpha^2}}\Big|
\end{eqnarray}
The Bell-Mermin function $BM_{\mathcal{A}}$
has been plotted in Fig.~\ref{fig:wbell2}. The Bell-Mermin
inequality begins to be violated from $\alpha \approx 1.49$ and the
value of $BM_{\mathcal{A}}$ rapidly saturates to $3$ as $\alpha $ grows.

\section{Conclusion}

We have studied Mermin's version of the Bell inequality for $GHZ$-type and 
$W$-type three-mode ECSs. Both the types of ESCs violate the Bell-Mermin
inequality with threshold photon detection (i.e., without photon counting). 
Such an experiment can be performed using linear optics elements and threshold
detectors with large violations for $GHZ$-type ESCs. However, it would be
experimentally more difficult to demonstrate Bell-type inequality violations
for $W$-type ECSs since it requires additional strong nonlinear
interactions. We have found out that the photon threshold measurements are
more efficient to test the Bell-Mermin inequality for the 
$GHZ$-type ECSs of small amplitudes, but for the $W$-type ECSs of large
amplitudes: an interesting fact that reflects a clear inequivalence
between the two types of ECSs. 
The generation of a $GHZ$-type ECS requires a CSS and two beam splitters.  
Recently, there was an experimental report to generate a CSS in a real laboratory
even though the fidelity was limited \cite{Wenger}.
Based on the recent theoretical \cite{Martini,Dakna2,Mauro03,Lund04,Jeong04,Jeong05,Wang,KimPaternostro05}
and experimental \cite{Wenger} progress,
it is expected that the realization of a CSS with higher fidelity
can be achieved in the foreseeable future.

We have also proposed, for the first time,  
a scheme to generate $W$-type ECSs for three
free-traveling optical fields (generalization to more than three fields is
possible). 
%The required resources are an imperfect single-photon
%source, a coherent state source, beam splitters, phase shifters and
%imperfect photodetectors and weak nonlinearities.
The required resources are
a %imperfect 
single-photon source, a coherent state source, beam splitters,
phase shifters, photodetectors, and 
%weak 
Kerr nonlinearities.
%%A perfect single-photon source and strong nonlinearities are not necessary.
%Our scheme does not necessarily
%require strong nonlinear interactions.
%%,
%%and the single-photon source and photodetectors do not
%%have to be highly efficient.
%It is also robust
%against inefficiencies of the single-photon source and the photon
%detectors.
Our scheme does not necessarily
require strong Kerr nonlinear interactions, i.e., weak nonlinearities
can be useful for our scheme.
Furthermore, it is also robust
against inefficiencies of the single-photon source and the photon
detectors.

 It should be noted that good mode matching would be required at
BS5 and BS6 in Fig.~\ref{fig:produce}, while efficient mode matching between
optical fields at a beam splitter is being performed in a real laboratory 
condition using present technology \cite{Pitt}. The dark
count rate of photodetectors will affect the fidelity of the generated
$W$-type ECSs. Currently, highly efficient detectors have relatively high dark
count rates while less efficient detectors have very low dark count rates 
\cite{Takeuchi}. 
We emphasize that our scheme here does not require
highly efficient detectors. 
Compared with the scheme to generate a CSS with
only one weak nonlinear interaction \cite{Jeong05}, from which the 
generation of $GHZ$-type ECSs can be straightforwardly performed, the most
demanding element in our scheme would be to control the three weak nonlinear
interactions in Fig~\ref{fig:produce}.
However, such techniques using weak
nonlinearities are being intensively studied 
\cite{Jeong05,KimPaternostro05,WKeffect,NM04}.

\acknowledgements
H.J. is supported by the Australian Research Council
while N.B.A. by the Vietnam Institute of Physics and Electronics
and Korea Institute for Advanced Study.

\end{document}